\documentclass[prl,aps,superscriptaddress,reprint]{revtex4-1}
\usepackage{graphicx} 
\usepackage{subfigure} %
\usepackage{dcolumn}
\usepackage{bm}
\usepackage{tabularx}
\usepackage[latin1]{inputenc}
\usepackage[usenames,dvipsnames]{color}
\usepackage[mediumspace,mediumqspace,Grey,squaren]{SIunits} 
\usepackage[version=3]{mhchem} 
\usepackage{amssymb} 

\def\be{\begin{equation}}
\def\ee{\end{equation}}
\def\ba{\begin{array}}
\def\ea{\end{array}}
\def\bea{\begin{eqnarray}}
\def\eea{\end{eqnarray}}

\begin{document}

\title{Fragmentation and limits to dynamical scaling  in viscous coarsening:\\
An interrupted in-situ X-ray tomographic study}

\date{\today}

\author{David Bouttes}
\affiliation{Laboratoire PMMH, UMR 7636 CNRS/ESPCI/Univ. Paris 6 UPMC/Univ. Paris 7 Diderot, 10 rue Vauquelin, 75231 Paris cedex 05, France}

\author{Emmanuelle Gouillart}
\affiliation{Surface du Verre et Interfaces, UMR 125 CNRS/Saint-Gobain, 93303 Aubervilliers, France}

\author{Elodie Boller}
\affiliation{European Synchrotron Radiation Facility (ESRF), BP 220, 38043 Grenoble, France}
\author{Davy Dalmas}
\affiliation{Surface du Verre et Interfaces, UMR 125 CNRS/Saint-Gobain, 93303 Aubervilliers, France}

\author{Damien Vandembroucq}
\affiliation{Laboratoire PMMH, UMR 7636 CNRS/ESPCI/Univ. Paris 6 UPMC/Univ. Paris 7 Diderot, 10 rue Vauquelin, 75231 Paris cedex 05, France}

\begin{abstract}
X-Ray microtomography was used to follow the coarsening of the
structure of a ternary silicate glass experiencing phase separation in
the liquid state. The volumes, surfaces, mean and Gaussian curvatures
of the domains of minority phase were measured after reconstruction of
the 3D images and segmentation. A linear growth law of the
characteristic length scale $\ell \sim t$ was observed.  A
  detailed morphological study was performed. While dynamical scaling
  holds for most of the geometrical observables under study, a
  progressive departure from scaling invariance of the distributions
  of local curvatures was evidenced. The latter results from a gradual
  fragmentation of the structure in the less viscous phase that also
  leads to a power-law size distribution of isolated domains.
\end{abstract}

\maketitle


{\sc Introduction \---} The construction of a theoretical framework
describing the phase separation of binary
~liquids\cite{Bray1994,Onuki-book,Allen-book} has been fueled by
successive experimental developments. Up to the 90's, most
experimental observations were performed in the Fourier space, with
light or neutron scattering.  Model systems were immiscible
solutions~\cite{Chou1979, Wong1981}, and polymer
blends~\cite{Bates1989, Takenaka1992}, or
glasses~\cite{Mazurin1985}. These experiments confirmed that the
dynamical scaling assumption is relevant for these systems, i.e. it is
possible to rescale the structure factor by a unique length scale
$\ell$. This length increases with a power law: $\ell(t) \sim
t^\alpha$, $\alpha$ depending on the growth regime \cite{Bray2003}.
More recent numerical simulations confirmed this scaling behavior, and
provided some insights about the geometry in real space
\cite{Wagner1998, Gonzalez-Segredo2003,
  Ahmad-Das-PRE2012}. Meanwhile, access to direct space was
  made possible by new techniques of observation, such as scanning
laser confocal microscopy: curvatures could be measured on
phase-separated polymer blends \cite{Jinnai2000,
  Lopez-Barron2009}. These observations are especially relevant to
discuss the local mechanisms that govern the coarsening: pinch-off
was for instance observed in a colloidal glass
\cite{Aarts2005}. Recent predictions concerning statistical quantities
such as size or surface distributions of domains\cite{Sicilia2007,
  Sicilia2009} and numerical studies of aging in
  phase-separating
  fluids\cite{Ahmad-Corberi-PRE2012,Majumder-Das-PRL2013} also
motivate the observation in real space. 

The effect of the mobility on the morphology has recently
  raised a growing interest. Experiments on polymers revealed the
spectacular influence of visco-elasticity \cite{Tanaka2000}. Recent
work~\cite{Testard2013} evidenced a logarithmic (stress-assisted) growth
for a gas-glass phase separation. Numerical simulations with a
viscosity contrast\cite{Novik2000, Luo2004} showed a strong effect on
morphology. However, the influence of a sole viscosity contrast has not
yet been experimentally investigated in the hydrodynamical regime.

The development of X-ray microtomography provides an suitable tool to
explore phase separation in 3-D, with submicron spatial
resolution reached in Synchrotron facilities~\cite{Baruchel2006, Buffiere2010}. 
Phase-separated polymer blends were already observed with phase-contrast imaging
\cite{Pyun2007}.

Glass-forming liquids offer an interesting opportunity to study phase
separation~\cite{Mazurin1985}. A fast quench below the glass transition
temperature enables one to study the frozen structure in the solid state,
thereby extending the variety of available characterization techniques.
Transmission and scanning electron microscopy, Raman Scattering and
Atomic Force Microscopy have been used to characterize quantitatively
phase separation in
glasses~\cite{Dalmas2007,Schuller-JACS11,Chopinet-JACS13}. We focus here
on the late stage of spinodal decomposition after a deep quench into the
unstable region, that produces interconnected structures. After an
initial stage where the interfaces form, the growth can be described by a
competition between the interfacial tension that favors the decrease of
the surface area between each phase, and dissipative forces (viscosity,
inertia).

In the following, we present first results on the observation of the
coarsening in silicate melts at high temperature, using X-ray
microtomography. After a description of the experimental system, a 3D
analysis of the main features of the coarsening stage is given. The
dynamical scaling hypothesis is tested up to its limits on a full set of
morphological observables. Associated with the onset of a fragmentation
process, a gradual departure from scaling invariance is observed on local
curvatures.


{\sc Materials and Methods \---} {\it A model glass for phase separation
\---} Using the knowledge available in glass science \cite{Mazurin1985},
we designed a barium borosilicate glass to study phase separation. The
composition of the glass was 57.1 \%wt \ce{SiO2}, 23.3 \%wt \ce{BaO}, 18.9
\%wt \ce{B2O3}. Elaborated from raw materials, this composition was
subsequently checked by wet chemical analysis. The glass decomposes
into two phases: a minority barium-rich one, and a majority barium-poor. This
 ensures a good absorption contrast for X-ray tomography, as barium is
much more absorbing than the other elements. 
This system has
a large metastable region which extends well above the liquidus
\cite{Levin1953, Levin1958}, which is appropriate to study the
coarsening of a binary liquid. 
The viscosity contrast in the range of temperatures of our experiments
is very high, the minority phase being much less viscous than the
majority phase. 

After elaboration, an interconnected microstructure of
typical size $\approx$ \unit{1}{\micro}{\meter} was already  present,
because of the large temperature
difference between the decomposition dome, and the glass transition
temperature, that permitted phase separation even during a fast quench.


{\it X-ray tomography \---} A series of microtomography experiments were
performed on the ID19 beamline at the European Synchrotron Radiation
Facility (ESRF). \unit{2}-{\milli}{\meter} diameter samples were placed
into refractory crucibles, the glass samples were then observed during a
heat treatment at a temperature corresponding to a liquid state. Three
experiments were done, at different temperatures: \unit{1080}{\celsius},
\unit{1130}{\celsius} and \unit{1180}{\celsius}, using an interrupted
in-situ protocol~\cite{Buffiere2010}. The samples were quenched in air at
regular time intervals to room temperature. They were subsequently
scanned, then heated again to the working temperature until the next
quench. The typical time scale for the temperature changes is of the
order of a few seconds. Because of the glassy nature of the material and
the relatively fast quench compared to the characteristic time of the
domain growth, it is expected that the successive quenches should not
influence the main features of the coarsening. We used a pink beam X-ray
radiation with a peak photon energy at 31 keV. For each scan, 900 X-ray
radiographies were recorded, from which the 3-D absorption field was
reconstructed using the standard filtered-back projection algorithm,
resulting in volumes of $512\times1024\times1024$ pixels. Pixel size was
\unit{0.7}{\micro\meter}, and scan time was approximately
\unit{5}{\minute}.


{\it Image processing \---} In order to study the geometrical
and temporal evolution of each phase during demixtion,
the following image processing was used.
First, reconstructed 3-D images were
denoised using a Total Variation filter
\cite{Chambolle2004}; the barium-rich and silica-rich phases were
subsequently identified with Random Walker algorithm
\cite{Grady2006} (see supplementary materials).
The different domains could then be identified as
distinct connected components. 
Then, geometrical characteristics, such as
volume $V_{i}$, surface $S_{i}$ of all domains $i$, and local curvatures, were computed
from the statistics of neighboring
configurations of 8-voxel cubes~\cite{Lang2001, Vogel2010}. Curvatures
were obtained from a local fit of the interface by a quadric~\cite{Sander1990}.

\begin{figure}[htp]
  \includegraphics[width=0.99\columnwidth]{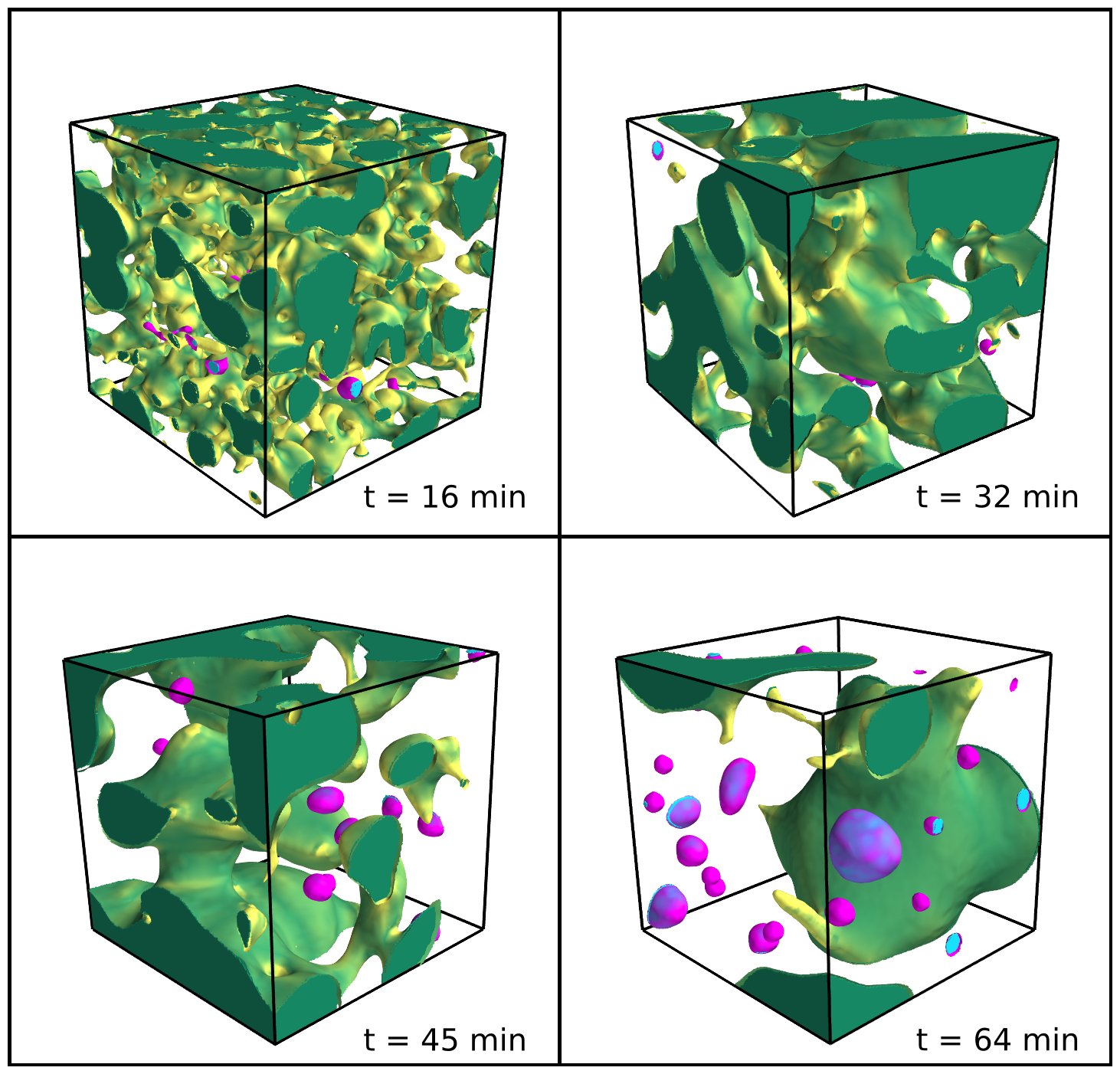}
  \caption{ 3D visualization\cite{ramachandran2011mayavi} of the
    barium-rich phase, from scans after different times at
    \unit{1130}{\celsius} (16, 32, 45
    and 64 min). The color codes represent the absolute value of the mean
    curvature: shades of green for the percolating domain,
    shades of purple for isolated domains. The lateral size of the cube is 
    \unit{140}{\micro}{\meter}. The coarsening of the structure is accompanied by a
    fragmentation of the spanning cluster into isolated domains.}
  \label{coarsening3D}
\end{figure}

{\sc Results \---} After reconstruction and segmentation, the
visual observation of a timeseries of images 
(Fig.~\ref{coarsening3D}) offers an intuitive idea of the coarsening.
Most of the minority phase
(barium rich), whose volume fraction is  $35 \pm 5 \%$, lies in a 
percolating network, but we observe also a growing fraction of
isolated domains when time increases. The coarsening appears clearly:
the width of the domains grows, and their surface gets smoother and
smoother.

In the following, we first study quantities averaged over the whole
volume, in analogy to previous experiments, to show the linear growth
in time of the typical scale. Then, we give a more detailed account of
the morphological features of the interconnected structure and their
time evolution. Finally, we characterize the development of a growing
fraction of isolated domains.


{\it Viscous coarsening \---} As mentioned before, various coarsening
regimes have been proposed for phase separation, depending on the
leading driving forces (diffusion, viscosity, inertia, etc.). The
general scaling law is $\ell \sim t^{\alpha}$. Here, we define $\ell$, the
characteristic length scale, as three times the ratio between 
the total volume $V_{tot}$ and the total
surface area $S_{tot}$ of the minority phase:
\begin{equation}
 \ell = 3\frac{V_{tot}}{S_{tot}}.
\end{equation}
The prefactor $3$ is chosen so that the characteristic length of a
system of spheres of radius $r = \ell$ is exactly $\ell$.

\begin{figure}[htp]
  \includegraphics[width=0.9\columnwidth]{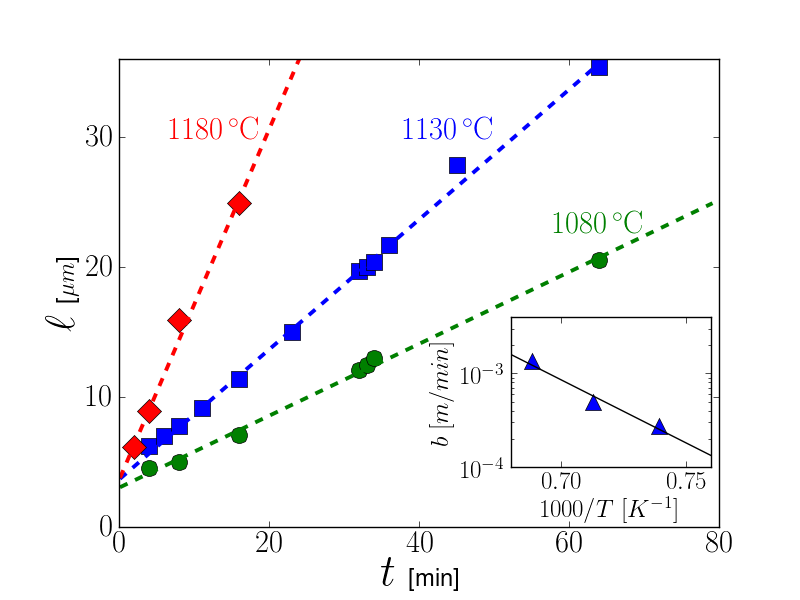}
  \caption{Characteristic length $\ell$ vs time $t$, for
    temperatures: \unit{1080}{\celsius} ({\Large\textcolor{OliveGreen}{$\bullet$}}),
    \unit{1130}{\celsius} (\textcolor{blue}{$\blacksquare$}), \unit{1180}{\celsius} 
  (\textcolor{red}{$\blacklozenge$}). The dotted lines are linear
    fits. Inset: semi-log plot of the slopes of the linear fits $b$,
    as a function of the inverse temperature $1/T$.}
  \label{lengths}
\end{figure}

The time evolution of this length is plotted on Fig.~\ref{lengths} for
T=\unit{1080}{\celsius}, \unit{1130}{\celsius} and
\unit{1180}{\celsius}, along with linear fits. As there was already a
microstructure at the beginning of the experiment, $\ell(t=0)\neq 0$. Our
results are consistent with the linear growth proposed by Siggia
\cite{Siggia1979} due to a hydrodynamic flow controlled by a
competition between surface tension $\gamma$ and viscosity $\mu$:

\begin{equation}
  \ell(t) \sim \frac{\gamma}{\mu} t.
\end{equation}

This mechanism assumes that the Laplace pressure due to local
curvatures induces a flow. It
requires that the minority phase is interconnected, since the flow inside
isolated domains stops as they approach a sphere. During the experiment, most of the minority phase remains in the
percolating domain (final volume ratio of about $75 \%$ to $
90\%$ in our experiments), so we do not expect a significant
slowing down on the measure of $\ell$. The coarsening is faster at higher
temperature, due to the decrease of viscosity. Growth rates at
different temperatures seem consistent with an Arrhenius law (inset, 
Fig. \ref{lengths}).

{\it Domain shape \---}
Using the data from individual domains, we characterize their shape from the
relation between their surface area and their volume.
In order to characterize the shapes at different times, we normalize the 
volume $V$ and the surface $S$ of each domain, using the characteristic length scale 
$\ell$:
\begin{equation}
 v = \frac{V}{\frac{4}{3}\pi \ell^{3}}
\quad ; \qquad
 s = \frac{S}{4\pi \ell^{2}}.
\end{equation}

\begin{figure}[htp]
\includegraphics[width=0.9\columnwidth]{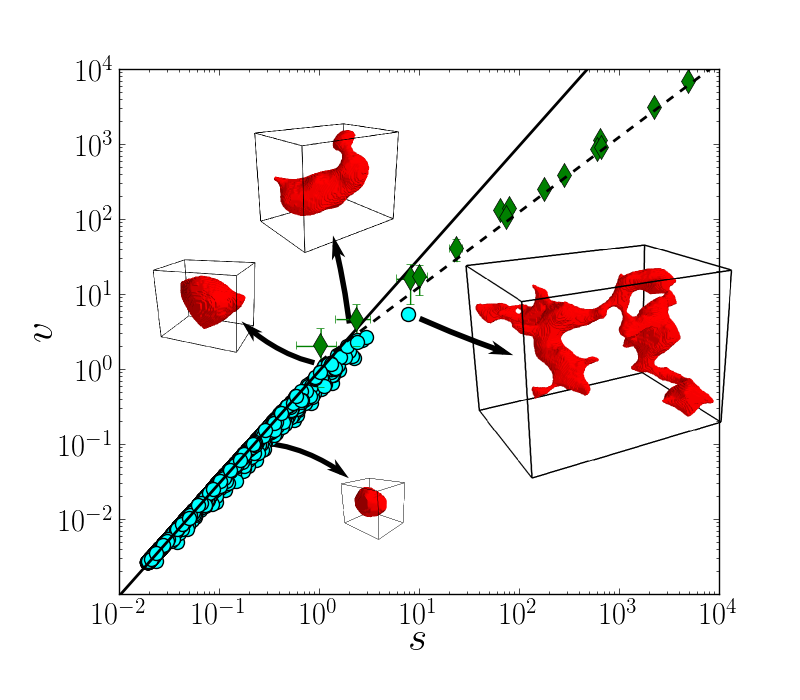}
  \caption{    Rescaled volumes $v$ of the domains vs their 
    rescaled surfaces $s$, for 
    all isolated domains (\textcolor{cyan}{{\Large$\bullet$}}) and for the
    percolating domain (\textcolor{OliveGreen}{$\blacklozenge$}) after 16, 32, 45 and \unit{64}{\minute} at \unit{1130}{\celsius}.
    The error bars indicate the standard deviation of the measures for all boxes of 
    a given size.
    The black solid line is $v = s^{3/2}$ (ideal spheres), and the black dotted line is $v \sim s$.
    Some domains shapes are shown as examples.}
  \label{surface_volume}
\end{figure}

The reduced volume $v$ is plotted as a function of the reduced volume
$s$ for all domains and different times in
Fig.~\ref{surface_volume}. When close to a sphere, the
volume of a domain should lie close to the solid line representing the
sphere case.  If  more elongated, or ramified (see
examples of domains in Fig.~\ref{surface_volume}), it should have an
excess of surface compared to a sphere, and hence get away from this
line.  According to this classification, we observe that most of the
small domains are close to a spherical shape, whereas bigger domains
deviate more and more as their volume increases. The dynamic
lengthscale $\ell(t)$ separates a dominant population of small
spheroidal clusters from a lower fraction of large ramified clusters.

Despite poor statistics on large domains, a natural assumption
would be to consider that the shape of large domains is inherited from
the percolating structure, as isolated domains detached from it.  They
would subsequently change their shape towards spheres and/or fragment
in more smaller domains.  If we plot the normalized volume of the
percolating domain $v_{p}$ as a function of its rescaled surface
$s_{p}$, measured at different times and in boxes of different sizes
within a given image, we find a linear relation: $v_{p} \sim s_{p}$,
see Fig~\ref{surface_volume}, dotted line. The large ramified domain
shape follows also the same linear relation, and the crossover between
the large and small domains corresponds well to the intersection of
the two asymptotic regimes.
Note that a non-trivial exponent could have been expected with an
initial condition closer to the percolation threshold in three
dimensions \cite{Sicilia2007, Sicilia2009}.

\begin{figure}[htp]
\includegraphics[width=0.99\columnwidth]{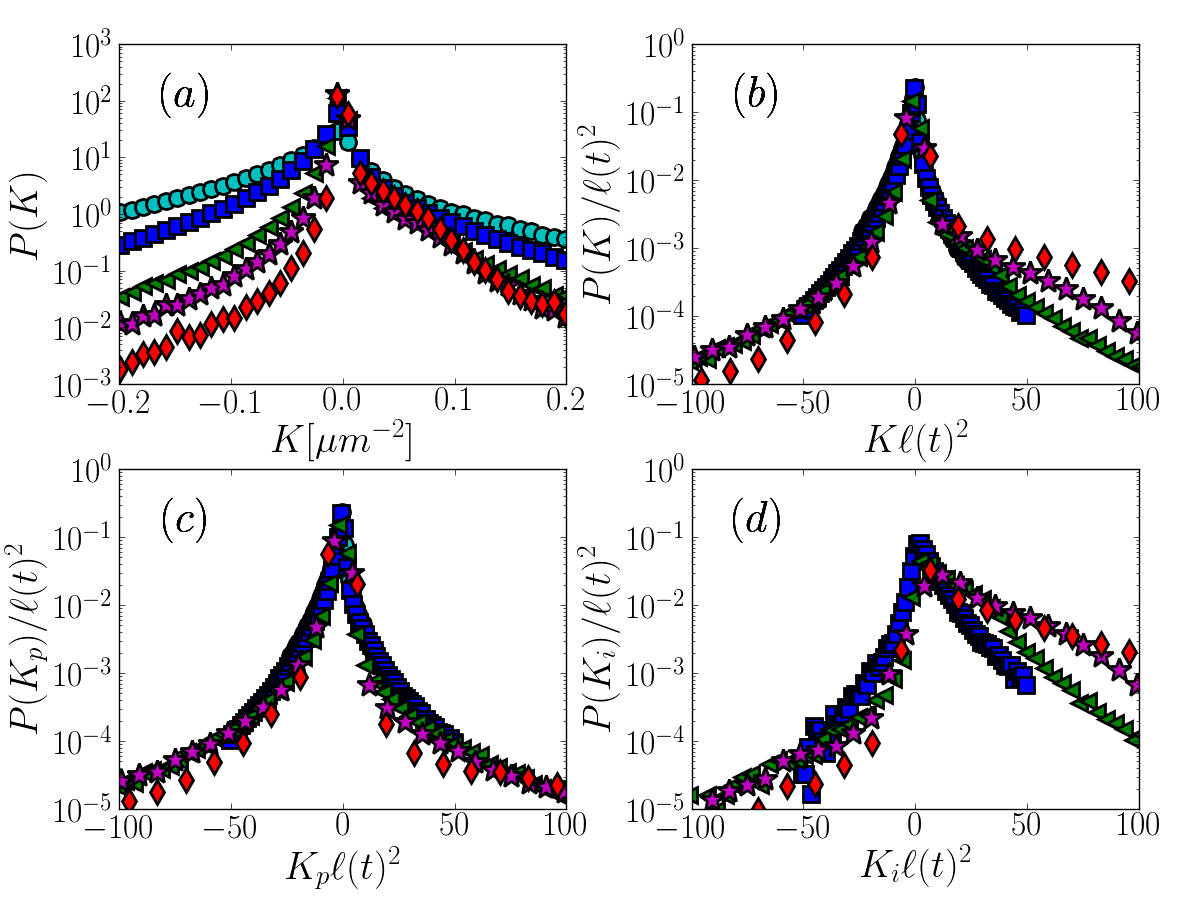}
  \caption{Gaussian curvature distributions $K$, $K_p$, $K_i$ ; before (a),
  and after rescaling for all domains (b), 
  only percolating (c) and only isolated domains (d)
  . After 8 (\textcolor{cyan}{{\Large$\bullet$}}), 
  16 (\textcolor{blue}{$\blacksquare$}),
  32 (\textcolor{OliveGreen}{$\blacktriangleleft$}), 
  45 (\textcolor{magenta}{$\bigstar$}) and \unit{64}{\minute} (\textcolor{red}{$\blacklozenge$})
  at \unit{1130}{\celsius}.}
  \label{Curvatures}
\end{figure}

{\it Local curvatures \---} Another benefit of real space imaging is
the possibility to access local geometrical observables such as local
curvatures~\cite{Lopez-Barron2009,Lopez-Barron2010} of the minority
phase.  In Fig.~\ref{Curvatures} (a) we show the distributions of the
Gaussian curvature $K$ obtained after 16, 23, 32, 45 and
\unit{64}{\minute} at \unit{1130}{\celsius}. As expected, the
distribution get narrower and narrower along the coarsening
process. The asymmetry of Gaussian curvatures reflects the topology of
the structure, the surface being dominated by
saddle-like shapes (negative value). For the whole system,
the time evolution appears to be reasonably well
captured by means of a simple rescaling by the dynamic length:
%
\[
   P(K) = \ell^2 \Psi\left[ K \ell(t)^2 \right]
\]
%
After rescaling (see Fig.~\ref{Curvatures} (b)) the distribution collapses
onto a master curve ($\Psi$), a similar
collapse is observed for the mean curvature (see supplementary
materials). Still, this collapse is limited by the fragmentation
of the percolating domain: an additional positive Gaussian curvature
appears with time due to the increasing number of isolated domains. 
Real-space imaging therefore 
enables us to show the origin of this gradual departure
from scaling invariance. When restricted to the percolating cluster 
($K_p$ Fig.~\ref{Curvatures} (c)),
a perfect collapse is recovered while a clear breaking of scaling
  invariance is obtained when the same analysis is performed within
  the sub-set of isolated domains  ($K_i$ Fig.~\ref{Curvatures} (d)).
  The collapse
  indicates that any measure of $\ell$ would be equivalent
  (which was checked, e.g. using chord length distribution, or
  the spatial correlation function). Besides, as only the
  percolating domain has a self-similar growth, one should 
  prefer a measure of $\ell$ that excludes other domains ;
  here their contribution is limited by their small 
  volume fraction.

\begin{figure}[htp]
  \includegraphics[width=0.9\columnwidth]{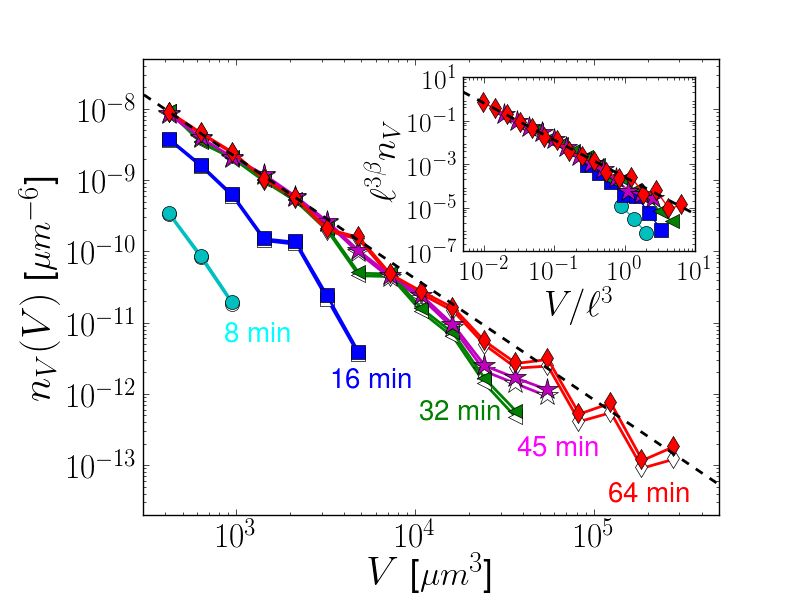}
  \caption{Domain volume distribution $n_{v}(V)$ 
    (except percolating domain)  after thermal
    treatment at T=\unit{1130}{\celsius} during 8, 16, 32, 45 and 64
    mn. The dotted line is $n_{V} \sim V^{-\beta}$, with $\beta =
    1.7$. The open symbols show the size distributions before the
    correction due to the finite size of the field of view. The inset
    shows the same results after rescaling by the dynamical length
    $\ell(t)$.}
  \label{size_distrib_1130}
\end{figure}

{\it Fragmentation \---} As was shown
previously in Fig~\ref{coarsening3D}, isolated domains are not
observed in the early stage of the experiment but they grow in number
during the experiment, as a consequence of
a fragmentation of the initial percolating structure. We show in
  Fig.~\ref{size_distrib_1130} the time evolution of size distribution
  of the domains per unit volume $n_{V}$, computed as the probability
  density
  to find a domain of volume $V$ per unit volume,
  measured at \unit{1130}{\celsius}.  Very small domains of linear size
smaller than 10 pixels are hard to identify with certainty and are
therefore not shown here.
  We computed
  the sampling bias affecting large domains touching the
  boundaries of the field of view (see supplementary material).
We observe that more domains are present at longer times. In
addition, the tail of the distribution changes with
time, since large domains are observed at longer times. Despite the
difficulty to obtain robust statistics for these large domains, the distribution seems to converge
towards a power law distribution: $\; n_{V} \sim V^{-\beta}$
with an exponent $\beta \approx 1.7 \pm 0.1$.

The power law distribution seems to develop due to the formation of
larger domains that detach from the percolating domain 
after some time, while the distribution of
small domains remains fairly constant once they are formed. This
suggests that an intrinsic size distribution exists and can be
rescaled by $\ell(t)$. The difficulty to sample very
small domains would hide the power law behavior for these domains at
early times. Hence we tested the scaling (with $\ell(t)^3$ as
characteristic volume):
\begin{equation}
 n_{V}(V) \sim \ell^{-3\beta} \Phi\left[\frac{V}{\ell^3}\right],
\end{equation}

and obtained a reasonable collapse of the data as illustrated in
Fig. \ref{size_distrib_1130} for the experiment performed at
\unit{1130}{\celsius}. 

With the assumptions that domains of volume $V$
detach at time $t \propto \ell(t) = V^{1/3}$, and that a larger number
$N_\text{break}$ of breaking events occur in a more ramified domain
according to $N_\text{break} \sim \ell^{-3}$, we obtain the distribution
of domain volumes $n_V = N_\text{break} \frac{dt}{dV} \sim V^{-5/3}$,
consistent with our observation $\beta \simeq 1.7$.


{\sc Conclusion \---} The use of interrupted in-situ X-ray tomography
gives access to a full 3-D characterization of the coarsening
process. Dynamic scaling was shown to account well for the evolution of all
geometrical observables. However, a detailed morphological
  analysis, including the measurements of local curvatures, allowed us
  to observe a progressive departure from scaling invariance. The
  latter results from the gradual fragmentation of the
  (minority) interconnected domain.  This process induces a complex
  multiscale microstructure, with a power-law distribution of isolated
  domain volumes, co-existing with the percolating phase.


\begin{acknowledgments}
{\sc Acknowledgments} This work was supported by the ANR program
``EDDAM" (ANR-11-BS09-027). Experiments were performed on beamline ID19 at
ESRF in the framework of proposal HD501. We gratefully acknowledge the
help of J. Grynberg, A. Lelarge, S. Patinet, F. Lechenault and
J.-P. Valade for performing the experiments at the ESRF, as well as
fruitful conversations with M.-H. Chopinet and L. Cugliandolo. We
thank Yohann Bale for his help in sample preparation and Erick Lamotte
for the glass elaboration.
\end{acknowledgments}

\bibliography{quenched}

\end{document}